\documentclass[12pt]{article}

\baselineskip=\normalbaselineskip

\usepackage{bm}
\usepackage{amsmath,amssymb,bm}
\usepackage{color}
\usepackage{cases}
\usepackage[dvipdfmx,colorlinks=true,linkcolor=blue,filecolor=blue,urlcolor=blue,citecolor=black]{hyperref}

\makeatletter
\@addtoreset{equation}{section}

\makeatother

\newcommand{\mr}[1]{\mathrm{#1}}
\newcommand{\delu}{\partial}
\newcommand{\nb}{\nabla}
\newcommand{\gm}{\Gamma}

\newcommand{\Lie}{\delta}

\newcommand{\shifthat}{{\vec{\hat{N}}}}
\newcommand{\hh}{{\mathsf{h}}}

\begin{document}

\begin{titlepage}
\renewcommand{\thefootnote}{\fnsymbol{footnote}}
\begin{flushright}
KUNS-2625
\end{flushright}
\vspace*{1.0cm}

\begin{center}
{\Large \bf 
Massive higher spin fields in curved spacetime\\ 
and necessity of non-minimal couplings
}
\vspace{1.0cm}

\centerline{
{Masafumi Fukuma,}%
\footnote{E-mail address: 
fukuma@gauge.scphys.kyoto-u.ac.jp} 
{Hikaru Kawai,}%
\footnote{E-mail address: 
hkawai@gauge.scphys.kyoto-u.ac.jp} 
{Katsuta Sakai}%
\footnote{E-mail address: 
katsutas@gauge.scphys.kyoto-u.ac.jp} and
{Junji Yamamoto}%
\footnote{E-mail address: 
junji@gauge.scphys.kyoto-u.ac.jp}%
}

\vskip 0.8cm
{\it Department of Physics, Kyoto University, Kyoto 606-8502, Japan}
\vskip 1.2cm 

\end{center}

\begin{abstract}

Free massive higher spin fields in weak background gravitational fields 
are discussed.
Contrary to the spin one case, 
higher spin fields should have nontrivial non-minimal couplings to the curvature. 
A precise analysis is given for the spin 2 case, 
and it is shown that two conditions should be satisfied 
among five non-minimal coupling constants, 
which we derive both in the Hamiltonian and Lagrangian formalisms. 
It is checked that the linearized limit of the massive gravity theory 
indeed has the non-minimal couplings that satisfy the conditions.
We also discuss the form of the non-minimal couplings for the spin 3 case. 

\end{abstract}
\end{titlepage}

\pagestyle{empty}
\pagestyle{plain}

\tableofcontents
\setcounter{footnote}{0}

\section{Introduction}
\label{introduction}

Attempts to construct massive higher spin field theories 
showed up with the papers written by Fierz and Pauli, 
who formulated a free field theory of massive spin 2 particles 
in the Minkowski space \cite{Fierz:1939zz}\cite{Fierz:1939ix}. 
In general, the natural object to describe a spin $s$ particle 
is a rank-$s$ traceless symmetric tensor field, 
but this has more independent components than necessary, 
because  a spin $s$ particle has only $2s+1$ degrees of freedom (DOF).
Therefore, the Lagrangian should give the equations of motion (EOM) 
that yield necessary and sufficient constraints 
to eliminate the redundant DOF. 
In fact, for the $s=2$ case, 
Fierz and Pauli showed that an appropriate Lagrangian can be obtained 
if one introduces an auxiliary scalar field 
in addition to a rank-2 traceless tensor. 
These fields can actually be combined 
to form a single traceful symmetric tensor $h_{\mu\nu}$, 
which we call the Fierz-Pauli (FP) field.%
\footnote{
 In \cite{VanNieuwenhuizen:1973fi} 
 it was shown that the FP theory is the unique formulation 
 of a spin 2 particle without ghosts or tachyons.
} 
For the case $s>2$, 
the Lagrangian with the desired property was given by Singh-Hagen 
\cite{Singh:1974qz}\cite{Singh:1974rc}, 
which consist of traceless symmetric tensors 
of ranks $s,\, s-2,\, s-3,\, s-4\,, \ldots,\, 0$\,. 
These fields can be combined to form two traceful symmetric tensors 
of ranks $s$ and $s-3$.%
\footnote{
 The massless limit of that Lagrangian was studied by Fronsdal 
 \cite{Fronsdal:1978rb}\cite{Fang:1978wz}.
} 

All the works above only consider the case 
where the background spacetime is flat. 
However, for curved backgrounds, 
it is non-trivial to formulate massive higher spin field theories.%
\footnote{
 For specific types of background, 
 consistent EOM are obtained for massless fields 
 by using the spacetime symmetry \cite{Fradkin:1986ka}%
 \cite{Fradkin:1986qy}\cite{Fradkin:1987ks}\cite{Vasiliev:1990en}. 
 An attempt to generalize the theory to the massive case was made 
 in \cite{Zinoviev:2015wpa}.
} 
In fact, as we will see in section \ref{basicFP}, 
the mechanism to derive the constraints from the EOM breaks down 
because covariant derivatives do not commute with each other. 
There was also an argument 
that the transverse condition is not compatible 
with the wave equation for arbitrary backgrounds \cite{Cortese:2013lda}. 
It seems that currently there are no consistent massive higher spin theories 
for general backgrounds that reduces to the flat case smoothly.

On the other hand, we expect that such theories should exist 
for the following two reasons. 
One is that phenomenologically higher spin hadrons should exist 
in the gravitational field. 
The other is that string theory consistently contains higher spin modes 
interacting with gravitons. 
In this paper, as a first step to investigate higher spin theories, 
we give the quadratic Lagrangian for spin 2 particles 
in general gravitational backgrounds.

This paper is organized as follows. 
In section \ref{basicFP}, 
we first show that the mechanism to eliminate the redundant DOF 
in the flat spacetime no longer works for general curved backgrounds. 
Then in section \ref{canonical}, 
we give a consistent quadratic Lagrangian of the massive spin 2 field 
in general backgrounds. 
To do that, we use the fact 
that the kinetic term of the FP field can be identified 
with the quadratic part in the perturbed Einstein-Hilbert action 
around the background metric. 
The analysis is based on the Hamiltonian formalism with the ADM decomposition. 
We find that a consistent theory can be constructed 
only when non-minimally coupled curvature terms 
are added to the Lagrangian with specific coefficients. 
In section \ref{nmFP} we reproduce the conditions on the coefficients 
within the Lagrangian formalism. 
In section \ref{spin3} we apply our analysis to the spin 3 case 
and investigate the form of the non-minimal couplings. 
Finally in section \ref{discussion}, 
we discuss the relation between our results on the spin 2 case 
and the massive gravity theory 
\cite{deRham:2010ik}\cite{deRham:2010kj}\cite{Hassan:2011vm}%
\cite{Hassan:2011tf}.


\noindent
{\bf [Note Added]}

After the first manuscript of this paper was accepted for publication, 
we were informed that the main result in section \ref{canonical} and \ref{nmFP}
were already obtained in \cite{Buchbinder:1999be}\cite{Buchbinder:1999ar}. 
We thank I.L.\ Buchbinder, M.\ von Strauss and A.\ Waldron 
for their valuable comments.
We were also informed of related works 
\cite{Buchbinder:2011vp}\cite{Bernard:2015mkk}\cite{Bernard:2015uic}\cite{Hassan:2012wr}\cite{Francia:2007ee}\cite{Francia:2008ac}\cite{Francia:2010ap}\cite{Hack:2011yv}\cite{Heisenberg:2014rta}\cite{Jimenez:2016isa}\cite{Cucchieri:1994tx}\cite{Deriglazov:2015zta}\cite{Deriglazov:2015wde}\cite{Germani:2004jf}\cite{Germani:2007em}. 

\section{Breakdown of the transverse condition for curved backgrounds}
\label{basicFP}

In this section, we demonstrate that FP's original mechanism 
to eliminate the redundant DOF of a massive higher rank tensor field  
does not work for generic curved backgrounds. 

We start by arguing that there is no such issue 
for massive spin 1 field $A^\mu$ (Proca field). 
The action of the Proca field in the flat Minkowski spacetime 
is given by 
\begin{align}
 S = \int d^4 x \,\Bigl[-\frac{1}{4}\,F^{\mu\nu} F_{\mu\nu}
 - \frac{1}{2}\,m^2\,A^\mu A_\mu \Bigr]\,,
\label{Proca_action_flat}
\end{align}
where $F_{\mu\nu}=\partial_\mu A_\nu - \partial_\nu A_\mu$ 
and the metric is chosen to be 
$\eta_{\mu\nu}={\rm diag}[-1,+1,+1,+1]$\,. 
Its EOM are given by 
\begin{align}
 \partial_\nu F^{\mu\nu} + m^2\,A^\mu = 0\,.
\label{Proca_EOM_flat}
\end{align}
The divergence of \eqref{Proca_EOM_flat} gives 
the transverse condition $\partial_\mu A^\mu=0$\,, 
and the substitution of this to the EOM 
in turn gives the wave equation, $(\Box-m^2)\,A^\mu=0$\,. 
Thus, the action \eqref{Proca_action_flat} gives the EOM 
which automatically include the constraint 
that eliminates the redundant DOF correctly. 
It is easy to see 
that this mechanism also works in general curved backgrounds. 
In fact, if we covariantize the action as  
\begin{align}
 S = \int d^4 x\,\sqrt{-g}\,
 \Bigl[ -\frac{1}{4}\,F^{\mu\nu} F_{\mu\nu}
 - \frac{1}{2}\,m^2\,A^\mu A_\mu \Bigr]
\end{align}
with $F_{\mu\nu}\equiv \nabla_\mu A_\nu - \nabla_\nu A_\mu$\,,
then the EOM are given by 
\begin{align}
 \nabla_\nu F^{\mu\nu} + m^2 A^\mu = 0\,, 
\end{align}
whose divergence again gives the transverse condition, 
$\nabla_\mu A^\mu = 0$\,,  
because $\nabla_\mu \nabla_\nu F^{\mu\nu} 
= [\nabla_\mu,\nabla_\nu]\,F^{\mu\nu}
= R_{\mu\nu}{}^\mu{}_\alpha F^{\alpha\nu}
+ R_{\mu\nu}{}^\nu{}_\alpha F^{\mu\alpha}
= -2 R_{\mu\nu} F^{\mu\nu} = 0$\,.%
\footnote{
The Riemann tensor is defined as 
$[\nabla_\mu,\nabla_\nu]\,v^\rho 
= R_{\mu\nu}{}^\rho{}_\sigma\,v^\sigma$\,. 
The Ricci tensor and the Ricci scalar are given by 
$R_{\mu\nu}\equiv R_{\rho \mu}{}^\rho{}_\nu$ 
and $R=g^{\mu\nu}\,R_{\mu\nu}$\,, respectively.  
} 
Note that one could have added  
curvature terms to the action
of the form $\int d^4 x\,\sqrt{-g}\,
\bigl[a\,R_{\mu\nu} A^\mu A^\nu + b\,R\,A^\mu A_\mu\bigr]$, 
where the coupling constants $a$ and $b$ are not determined 
only by requiring the action to become \eqref{Proca_action_flat} 
in the flat limit. 
Such non-minimal couplings can be used to absorb the discrepancy 
that may arise when kinematic terms are covariantized in a different manner 
[e.g., a kinetic term $\partial_\mu A^\mu\, \partial_\nu A^\nu$ 
(up to total derivatives) can be covariantized in two ways: 
$\nabla_\mu A^\mu\, \nabla_\nu A^\nu$ 
or $\nabla_\mu A^\nu\, \nabla_\nu A^\mu$].

Now we discuss the spin 2 massive field (FP field). 
The Lagrangian in the flat spacetime is given by
\begin{align}
 \mathcal{L} &= 
 h_{\mu\nu}\,\mathcal{E}_0^{\mu\nu\rho\sigma} h_{\rho\sigma}
 -\frac{m^2}{2}\,(h_{\mu \nu}h^{\mu \nu}-h^2)\,,
\label{fhlgn}
\end{align}
where $\mathcal{E}_0^{\mu\nu\rho\sigma}$ 
is the Lichnerowicz operator for the flat spacetime:%
\footnote{
We normalize the symmetrization as 
$X^{(\mu\nu)}\equiv (1/2)\,(X^{\mu\nu}+X^{\nu\mu})$.
} 
\begin{align}
 \mathcal{E}_0^{\mu\nu\rho\sigma} h_{\rho\sigma}
 \equiv
 \, \frac{1}{2}(\Box h^{\mu \nu} - \eta^{\mu\nu} \Box h)
 +\frac{1}{2}(\delu^\mu \delu^\nu h
 +\eta^{\mu \nu}\partial^\rho \partial^\sigma h_{\rho \sigma})
 -\partial^{(\mu} \partial_\lambda h^{\nu)\lambda}\,.
\label{Lichnerowicz_flat}
\end{align}
The kinetic term 
$\mathcal{L}_{\mathcal{E}_0}
=h_{\mu\nu}\,\mathcal{E}_0^{\mu\nu\rho\sigma} h_{\rho\sigma}$ 
can be formally obtained 
from the Einstein-Hilbert action%
\footnote{
Throughout this paper 
quantities with turret should be understood 
to represent those associated with $\hat{g}_{\mu \nu}$\,. 
} 
\begin{align}\label{EH}
 S_{\mr{EH}}[\hat{g}]=\frac{1}{2}\int d^4x\sqrt{-\hat{g}}\hat{R}
\end{align}
by setting $\hat{g}_{\mu \nu} = \eta_{\mu \nu}+2 h_{\mu \nu}$ 
and taking quadratic terms in $h_{\mu\nu}$\,.
The EOM take the form
\begin{align}
 0&={} 2\, \mathcal{E}_0^{\mu\nu\rho\sigma} h_{\rho\sigma}
 -m^2 (h^{\mu\nu}-\eta^{\mu\nu}\,h)
\nonumber
\\
 &=(\Box-m^2)(h^{\mu \nu}-\eta^{\mu \nu}h)
 +\eta^{\mu \nu}\delu^\rho \delu^\sigma h_{\rho \sigma}
 -2\delu^{(\mu}\delu_{\lambda}h^{\nu) \lambda}+\delu^\mu \delu^\nu h\,.
\label{fheom}
\end{align}
A rank-2 symmetric tensor $h_{\mu\nu}$ has ten independent components, 
while a massive spin 2 particle has five DOF. 
In the flat background, 
the extra DOF are actually eliminated from the EOM as the Proca field. 
In fact, the divergence, double divergence, 
and trace of (\ref{fheom}) respectively give
\begin{align}
 -m^2(\delu_\nu h^{\mu \nu}-\delu^\mu h) = 0 \,,   \label{fhd} \\
 -m^2(\delu_\mu \delu_\nu h^{\mu \nu}-\Box h) = 0 \, ,  \label{fhdd} \\ 
 2(\delu_\mu \delu_\nu h^{\mu \nu}-\Box h)+3m^2h = 0 \, .  \label{fhtr}
\end{align}
Thus, when $m\neq 0$, 
we obtain the traceless condition, $h=0$, from (\ref{fhdd}) and (\ref{fhtr}). 
Then, substituting it to (\ref{fhd}), 
we get the transverse condition, $\delu_\nu h^{\mu \nu}=0$. 
Consequently, $h_{\mu \nu}$ is a rank-2 traceless symmetric, 
divergence-free tensor, which has five independent components. 
Note that the EOM (\ref{fheom}) are then reduced to the Klein Gordon equations:
\begin{align}\label{fheom2}
 (\Box-m^2)h^{\mu \nu}= 0\, .
\end{align}
We thus see that the reduction mechanism works for a massive spin 2 field 
as long as the background is flat.

Next we show the breakdown of the reduction mechanism 
when the flat theory is na\"ively lifted to curved backgrounds. 
A natural extension of (\ref{fhlgn}) is obtained 
(a) by replacing the derivatives (\ref{fhlgn}) with covariant derivatives, 
or (b) by substituting $\hat{g}_{\mu \nu} = g_{\mu \nu} + 2h_{\mu \nu}$ 
to (\ref{EH}) 
and taking only quadratic terms in $h_{\mu \nu}$. 
The discrepancy between (a) and (b) appears 
as the difference of non-minimal couplings 
(e.g., the difference of the coefficient of $R\, h^{\mu\nu} h_{\mu\nu}$). 
In this section we adopt the prescription (b).

The Lagrangian now takes the form
\begin{align}
\mathcal{L}
 &=\sqrt{-g}\Bigl[h_{\mu \nu}\mathcal{E}^{\mu \nu \rho \sigma}h_{\rho \sigma}-\frac{m^2}{2}(h_{\mu \nu}h^{\mu \nu}-h^2)\Bigr] \,.
\label{Lagrangian_FP_curved}
\end{align}
Here, $h=g^{\mu\nu} h_{\mu\nu}$\,, 
and $\mathcal{E}^{\mu \nu \rho \sigma}$ is the Lichnerowicz operator 
acting on symmetric tensors in a curved spacetime:
\begin{align}
 \mathcal{E}^{\mu \nu \rho \sigma}h_{\rho \sigma} 
 &= \frac{1}{2}(\Box h^{\mu \nu}-g^{\mu \nu}\Box h)
 +\frac{1}{2}(\nb^\mu\nb^\nu h
 +g^{\mu \nu}\nb^\rho\nb^\sigma h_{\rho \sigma})
 -\nb^{(\mu}\nb_\lambda h^{\nu) \lambda} 
\nonumber
\\
 &~~~+R^{\mu \rho \nu \sigma}h_{\rho \sigma}+R^{\rho (\mu}h_\rho^{\nu)}
 -\frac{1}{2}(g^{\mu \nu}R^{\rho \sigma}h_{\rho \sigma}+R^{\mu \nu}h)
 -\frac{1}{2}Rh^{\mu \nu}+\frac{1}{4}Rg^{\mu \nu}h \,,
\label{Lichnerowicz} 
\end{align}
which reduces to \eqref{Lichnerowicz_flat} in the flat limit 
and enjoys the following properties: 
\begin{align}
 &\frac{1}{2}\,\sqrt{-\hat{g}}\,\hat{R}
 = \sqrt{-g}\,\Bigl[
 \frac{1}{2}\,R - G^{\mu\nu} h_{\mu\nu}
 + h_{\mu\nu}\,\mathcal{E}^{\mu\nu\rho\sigma} h_{\rho\sigma}
 + O(h^3)\Bigr]
 ~~~\bigl(\hat{g}_{\mu\nu}=g_{\mu\nu}+ 2 h_{\mu\nu}\bigr)\,,
\\
 &\nabla_\nu \bigl(\mathcal{E}^{\mu\nu\rho\sigma} h_{\rho\sigma} \bigr)
 =\frac{1}{2}\, G^{\rho \sigma}\left( 2\nb_\rho h_\sigma^\mu 
 - \nb^\mu h_{\rho  \sigma}\right)\,,
\\
 &g_{\mu\nu} \,\bigl(\mathcal{E}^{\mu\nu\rho\sigma} h_{\rho\sigma}\bigr)
 =\nb_\mu \nb_\nu h^{\mu \nu}-\Box\, h \,,
\end{align}
where $G^{\mu\nu}=R^{\mu\nu}-(R/2)\,g^{\mu\nu}$ is the Einstein tensor. 
The EOM are given by
\begin{align}
 2\mathcal{E}^{\mu \nu \rho \sigma}h_{\rho \sigma}
 -m^2(h^{\mu \nu}-g^{\mu \nu}h)=0 \label{cheom}\,.
\end{align}
The divergence, double divergence, and trace of (\ref{cheom}) 
respectively give
\begin{align}
 G^{\rho \sigma}\left( 2\nb_\rho h_\sigma^\mu 
 -\nb^\mu h_{\rho \sigma}\right)-m^2(\nb_\nu h^{\mu \nu}-\nb^\mu h) &= 0 \,, 
\label{chd} 
\\
 \nb_\mu \bigl[ G^{\rho \sigma}\left( 2\nb_\rho h_\sigma^\mu 
 - \nb^\mu h_{\rho  \sigma}\right)\bigr]
 -m^2(\nb_\mu \nb_\nu h^{\mu \nu}
 -\Box\, h) &= 0 \,,   
\label{chdd} 
\\
 2(\nb_\mu \nb_\nu h^{\mu \nu}-\Box\, h)+3m^2h &= 0 \,.  
\label{chtr}
\end{align}
Thus, if $h$ vanished or at least could be expressed 
as a function of the traceless part of $h_{\mu \nu}$, 
(\ref{chd}) would give four constraints 
on the transverse component.  
However, (\ref{chdd}) and (\ref{chtr}) lead to
\begin{align}
 h=-\frac{2}{3m^4}\nb_\mu 
 \bigl[ G^{\rho \sigma}\left( 2\nb_\rho h_\sigma^\mu
 - \nb^\mu h_{\rho \sigma}\right) \bigr] \,.
\label{chnotconst}\end{align}
This is, except for the vacuum case ($G_{\mu \nu}=0$), 
a second-order differential equation 
for the trace $h$ and the traceless part of $h_{\mu\nu}$, 
which cannot be regarded as a constraint eliminating unnecessary DOF. 
The situations are the same also for the cases of other spins, 
except for spin 1 (Proca field). 
In the spin $1$ case, the divergence of the EOM always results 
in a first-order differential equation 
corresponding to the transverse condition, 
irrespective of how non-minimal couplings are introduced. 
For the case of higher spins, however, 
there is no choice of non-minimal couplings 
so as to cancel the RHS of (\ref{chnotconst}). 
Another problem will emerge when formally substituting (\ref{chnotconst}) 
to (\ref{cheom}), 
since it results in fourth-order differential equations with respect to time. 
It is a singular perturbation, 
and yields an exponential growth of the amplitudes 
because the perturbation becomes much larger 
than the original kinetic term at short time scales. 
These facts seem to indicate that the Lagrangian above fails 
to describe a consistent FP field in a general background. 
In the following, we resolve this issue 
by giving up the attempt 
to express the constraint 
in a form that is directly related to the transverse condition  
and also by paying the cost of breaking the manifest covariance in the analysis.

\section{Fierz-Pauli field in general curved backgrounds}
\label{canonical}

In this section, we construct a consistent, linear field theory 
of a massive spin 2 field in a general curved spacetime.

We start with the Lagrangian (\ref{Lagrangian_FP_curved})
with non-minimal couplings to the curvature:
\begin{align}
 S&=\int d^4x \mathcal{L},\hspace*{12pt} \mathcal{L}
 = \mathcal{L}_\mathcal{E} + \mathcal{L}_m + \mathcal{L}_R
\label{non-min S}
\end{align}
with 
\begin{align}
 &\mathcal{L}_\mathcal{E}
 = \sqrt{-g}\, h_{\mu \nu} \mathcal{E}^{\mu \nu \rho \sigma}
 h_{\rho \sigma},\hspace*{12pt}
 \mathcal{L}_m = - \sqrt{-g}\, \frac{m^2}{2}{(}h_{\mu \nu} h^{\mu \nu} - h^2{)}\,,
\\
 &\mathcal{L}_R = \sqrt{-g}\, \Bigl[ \frac{a_1}{2} R_{\mu \nu \rho \sigma}
 h^{\mu \rho} h^{\nu \sigma} + \frac{a_2}{2} R_{\mu \nu} h^{\mu \rho}
 h^\nu_\rho + \frac{a_3}{2} R h_{\mu \nu} h^{\mu \nu} + \frac{b_1}{2} R h^2
 + b_2 R_{\mu \nu} h^{\mu \nu} h \Bigr]\,.
\label{R term}
\end{align}
Here $\mathcal{L}_R$ expresses the non-minimal couplings, and
the coupling constants $a_1$, $a_2$, $a_3$, $b_1$, $b_2$ 
cannot be determined {\it a priori} 
only by requiring the action to become the FP action in the flat limit. 
Note that such terms also exist in $\mathcal{L}_\mathcal{E}$. 
In the remaining of this section, 
we show that 
the action (\ref{non-min S})--(\ref{R term}) 
describe a massive spin 2 field 
with correct DOF 
if and only if the constants in $\mathcal{L}_R$ satisfy 
the two conditions%
\footnote{
 The relations were first obtained in \cite{Buchbinder:1999ar}.
} 
\begin{align}
 a_2 + 2b_2 = -1 \,,
\label{constraint1}\\
 a_3 + b_1 = \frac{1}{2} \,.
\label{constraint2}
\end{align}

The counting of DOF is usually easiest 
in the Hamiltonian formalism, 
and for this purpose we introduce the ADM decomposition of the metric: 
\begin{align}
 (\hat{g}_{\mu \nu})=\left(
 \begin{array}{cc}
 -\hat{N}^2 + \hat{g}_{ij} \hat{N}^i \hat{N}^j & \hat{g}_{ij} \hat{N}^i \\
 \hat{g}_{ij} \hat{N}^j & \hat{g}_{ij}
 \end{array}
 \right).
\end{align}
The functions $\hat{N}$ and $\shifthat=(\hat{N}^i)$ $(i=1,2,3)$ are called 
the lapse and the shift, respectively,  
and $\hat{g}_{ij}$ describes the induced metric on a timeslice. 
The Einstein-Hilbert action then takes the following form 
up to surface integrals: 
\begin{align}
 S_{EH} = \int d^4x \frac{1}{2} \hat{N} \sqrt{\hat{g}}\, \bigl{[} {}^{(3)}\!\hat{R}
 + \hat{K}_{ij}\hat{K}^{ij} - \hat{K}^2 \bigr{]}
 \quad
 (\hat{K} \equiv \hat{g}^{ij}\,\hat{K}_{ij})\,.
\label{EH_ADM}
\end{align}
Here, ${}^{(3)}\!\hat{R}$ is the Ricci scalar 
associated with $\hat{g}_{ij}$\,, 
and 
$\hat{K}_{ij}\equiv (1/2\hat{N})\, \bigl[\dot{\hat{g}}_{ij}
 - \Lie_{\shifthat} \,\hat{g}_{ij}\bigr]$ 
is the extrinsic curvature of the timeslice 
($\Lie_{\shifthat}$ is the Lie derivative 
with respect to the shift $\shifthat$). 
We now expand the action around a classical background metric. 
By using the diffeomorphism invariance of the Einstein-Hilbert action, 
we can set the background to the following form 
without loss of generality: 
\begin{align}
 \bigl(g_{\mu\nu}\bigr)
 =\left(
 \begin{array}{cc}
 -1  & 0 \\
 0 & g_{ij}
 \end{array}
 \right). 
\label{metric_temporal}
\end{align}
We then replace the metric in the action as  
\begin{align}
 \hat{g}_{\mu\nu} = g_{\mu\nu} + \hh_{\mu\nu}
 \quad(\hh_{\mu\nu}\equiv 2h_{\mu\nu})\,,
\end{align}
or equivalently, 
rewrite the lapse and shifts in (\ref{EH_ADM}) as 
\begin{align}
	\hat{N}^2 &= 1- \hh_{00} +  \hat{g}^{ij} \hh_{0i} \hh_{0j}\,,
\label{lapse}
\\
 \hat{g}_{ij} \hat{N}^j &= \hh_{0i}\,,
\label{shift}
\\
 \hat{g}_{ij} &= g_{ij} + \hh_{ij}\,.
\label{spatial metric}
\end{align}
The quadratic terms in $\hh_{\mu \nu}$ give $\mathcal{L}_{\mathcal{E}}$, 
whose explicit form is given by 
\begin{align}
 \mathcal{L}_{\mathcal{E}} &=\biggl{[} \frac{1}{2} \hat{N} \sqrt{\hat{g}}\,
 \bigl{[} {}^{(3)}\!\hat{R}
 + \hat{K}_{ij}\hat{K}^{ij} - \hat{K}^2 \bigr{]} \biggr{]}_{(2)} 
\nonumber
\\
 &= \biggl{[} \frac{\hat{N}\sqrt{\hat{g}}}{2} {}^{(3)}\!\hat{R} + \frac{1}{2} \hat{C}^{ijkl}
 \bigl(\dot{\hat{g}}_{ij} - \Lie_{\shifthat} \hat{g}_{ij} \bigr)
 \bigl(\dot{\hat{g}}_{kl} - \Lie_{\shifthat} \hat{g}_{kl} \bigr)\biggr{]}_{(2)} 
\nonumber
\\
 &= \frac{1}{2} \hat{C}_{(0)}^{ijkl} \dot{\hh}_{ij} \dot{\hh}_{kl}
 + \hat{C}_{(1)}^{ijkl} \dot{\hh}_{ij} \dot{g}_{kl}
 - \hat{C}_{(0)}^{ijkl} \dot{\hh}_{ij} (\Lie_{\shifthat} g_{kl})_{(1)} 
\nonumber
\\
 & \hspace*{24pt} + \Bigl[\frac{\hat{N}\sqrt{\hat{g}}}{2} {}^{(3)}\!\hat{R}
 + \frac{1}{2} \hat{C}^{ijkl}  \bigl(\dot{g}_{ij}
 - \Lie_{\shifthat} g_{ij} - \Lie_{\shifthat} \hh_{ij}\bigr)
 \bigl(\dot{g}_{kl}
 - \Lie_{\shifthat} g_{kl} - \Lie_{\shifthat} \hh_{kl} \bigr) \Bigr]_{(2)}\,,
\end{align}
where 
\begin{align}
  \ \hat{C}^{ijkl} \equiv \frac{\sqrt{\hat{g}}}{4\hat{N}}\,
 \Bigl[\frac{1}{2} \,(\hat{g}^{ik} \hat{g}^{jl} + \hat{g}^{il} \hat{g}^{jk})
 - \hat{g}^{ij} \hat{g}^{kl} \Bigr], 
\end{align}
and a subscript in parenthesis 
denotes the order in $\hh_{\mu\nu}$.

We now move on to the Hamiltonian formalism 
by making the Legendre transformation with respect to $\dot{\hh}_{ij}$. 
Since $\dot{\hh}_{ij}$ is contained only in $\mathcal{L}_{\mathcal{E}}$\,, 
the conjugate variable to $\hh_{ij}$ is given by 
\begin{align}
 \pi^{ij} &\equiv \frac{\partial \mathcal{L}}{\partial \dot{\hh}_{ij}}
 = \frac{\partial \mathcal{L}_{\mathcal{E}}}{\partial \dot{\hh}_{ij}}
\nonumber
\\
 &= \hat{C}_{(0)}^{ijkl} \dot{\hh}_{kl} + \hat{C}_{(1)}^{ijkl} \dot{g}_{kl}
 - \hat{C}_{(0)}^{ijkl} (\Lie_{\shifthat} g_{kl})_{(1)}\,,
\end{align}
which can be solved for $\dot{\hh}_{ij}$ as 
\begin{align}
 \dot{\hh}_{ij} &= (\hat{C}^{-1}_{(0)})_{ijkl} \bigl( \pi^{kl}
 - \hat{C}_{(1)}^{klmn} \dot{g}_{mn} + {\hat{C}_{(0)}}^{klmn}
 (\Lie_{\shifthat} g_{mn})_{(1)} \bigr)\,.
\end{align}
The Hamiltonian is then obtained as 
\begin{align}
 \mathcal{H} &= \pi^{ij} \dot{\hh}_{ij} - \mathcal{L}_{\mathcal{E}}
 -\mathcal{L}_m -\mathcal{L}_R 
\nonumber
\\
 &= \frac{1}{2} (\hat{C}^{-1}_{(0)})_{ijkl} \bigl{(} \pi^{ij}
 - \hat{C}_{(1)}^{ijmn} \dot{g}_{mn}
 + \hat{C}_{(0)}^{ijmn} (\Lie_{\shifthat} g_{mn})_{(1)} \bigr)
 \bigl( \pi^{kl} - \hat{C}_{(1)}^{klpq} \dot{g}_{pq} + \hat{C}_{(0)}^{klpq}
 (\Lie_{\shifthat} g_{pq})_{(1)} \bigr{)} 
\nonumber
\\
 & \hspace*{24pt} - \Bigl[ \frac{\hat{N}\sqrt{\hat{g}}}{2} \,{}^{(3)}\!\hat{R}
 + \frac{1}{2} \hat{C}^{ijkl} (\dot{g}_{ij} - \Lie_{\shifthat} g_{ij}
 - \Lie_{\shifthat} \hh_{ij})(\dot{g}_{kl} - \Lie_{\shifthat} g_{kl}
 - \Lie_{\shifthat} \hh_{kl}) \Bigr]_{(2)}
 -\mathcal{L}_m - \mathcal{L}_R\,.
\label{Hamiltonian}
\end{align}

Since $\hh_{0i}$ is generically quadratic and has no kinetic terms, 
the corresponding DOF will drop out from the system 
by solving the EOM for $\hh_{0i}$ 
and by substituting the obtained solution to the action. 
Then, if the resulting Hamiltonian has only linear terms in $\hh_{00}$, 
there will arise the primary constraint, 
from which will follow the secondary constraint 
as a condition for the primary constraint 
to be consistent under the time evolution. 
Furthermore, a further consistency condition will arise 
for the secondary constraint, 
which in turn will determine the form of $\hh_{00}$. 
Thus, 
if the Hamiltonian has only linear terms in $\hh_{00}$ 
after the elimination of $\hh_{0i}$\,, 
the variables $\hh_{00}$ and $\hh_{0i}$ will disappear from the system, 
leaving two constraints. 
This means that the system has ten $(=6+6-2)$ DOF, 
which agree with those of a massive spin 2 field. 
We are going to show that this is the case 
if and only if the conditions (\ref{constraint1}) and (\ref{constraint2}) 
are met.

There are actually two sources of $\hh_{00}^2$ terms. 
One is the $\hh_{00}^2$ terms that already exist in the Hamiltonian 
before solving the EOM for $\hh_{0i}$\,. 
The other is the $\hh_{00}^2$ terms 
that come out after $\hh_{0i}$ is eliminated from the Hamiltonian.

First we point out that the latter source is absent, 
noticing that the mass term $\mathcal{L}_m$, 
\begin{align}
 \mathcal{L}_m = -\sqrt{-g} \,\frac{m^2}{8}
 \bigl[ -2g^{ij} \hh_{0i} \hh_{0j}
 + g^{ik} g^{jl} \hh_{ij} \hh_{kl}
 + 2 \hh_{00} g^{ij} \hh_{ij} - (g^{ij} \hh_{ij})^2
 \bigr]\,,
\label{mass term}
\end{align}
contains quadratic terms in $\hh_{0i}$ when $m\neq 0$\,.
If the Lagrangian contains the terms of the form $\hh_{00} \hh_{0i}$\,, 
the EOM for $\hh_{0i}$ 
take the form $\hh_{0i} = \hh_{00}\times A_{0i} + \cdots$  
and give $\hh_{00}^2$ terms when substituted back to the Lagrangian. 
However, as we will see below, 
there are no such terms in the Lagrangian. 
Since there are no $\hh_{00} \hh_{0i}$ terms in $\mathcal{L}_{m}$\,, 
we only need to confirm the absence of such terms 
in the rest of the Hamiltonian \eqref{Hamiltonian}. 
As for $\mathcal{L}_R$\,, 
we see that 
$a_2 R_{\mu \nu} \hh^{\mu \rho} \hh^\nu_\rho$ and 
$b_2 R_{\mu \nu} \hh^{\mu \nu} \hh$ 
actually give dangerous terms 
$-a_2 R_{0i} \hh_{00} \hh^i_0$ and $-2b_2 R_{0i} \hh_0^i \hh_{00}$.
However, they can be ignored in our present approximation, 
because their contributions to the coefficients of $\hh_{00}^2$ 
will be $O(R^2/m^2)$ 
and can be neglected to the first order in the curvature.  
As for the remaining part of \eqref{Hamiltonian}, 
we see from (\ref{lapse})--(\ref{spatial metric}) 
that terms linear in $\hh_{0i}$ appear 
only through $\Lie_{\shifthat} g_{ij}$\,. 
Thus, the possible terms containing $\hh_{00} \hh_{0i}$ are 
\begin{align}
 -(\hat{C}^{-1}_{(0)})_{ijkl} \hat{C}_{(1)}^{ijmn} 
 \dot{g}_{mn} \hat{C}_{(0)}^{klpq}
 (\Lie_{\shifthat} g_{pq})_{(1)} - \bigl{(}- \hat{C}^{ijkl} \dot{g}_{ij}
 \Lie_{\shifthat} g_{kl} \bigr{)}_{(2)}\,.
\label{cancel}
\end{align}
However, the $\hh_{00}\,\hh_{0i}$ terms cancel out in \eqref{cancel}, 
because it can be rewritten as 
\begin{align}
 & - \hat{C}_{(1)}^{ijmn} \dot{g}_{mn} (\Lie_{\shifthat} g_{ij})_{(1)} + \hat{C}_{(0)}^{ijkl} \dot{g}_{ij}
 \bigl{(} \Lie_{\shifthat} g_{kl} \bigr{)}_{(2)}  + \hat{C}_{(1)}^{ijkl} \dot{g}_{ij}
 \bigl{(} \Lie_{\shifthat} g_{kl} \bigr{)}_{(1)}  + \hat{C}_{(2)}^{ijkl} \dot{g}_{ij}
 \bigl{(} \Lie_{\shifthat} g_{kl} \bigr{)}_{(0)} 
\nonumber
\\
  &= \hat{C}_{(0)}^{ijkl} \dot{g}_{ij}
 \bigl{(} \Lie_{\shifthat} g_{kl} \bigr{)}_{(2)} + \hat{C}_{(2)}^{ijkl} \dot{g}_{ij}
 \bigl{(} \Lie_{\shifthat} g_{kl} \bigr{)}_{(0)} \,,
\end{align}
which does not contain $\hh_{00}\,\hh_{0i}$\,.

We thus find that $\hh_{0i}$ do not play any role 
in investigating the possible appearance of $\hh_{00}^2$ terms, 
so that we can safely set $\hh_{0i}=0$ for further arguments. 
Since $\hh_{00}^2$ terms can appear only through $\hat{N}$ in $\hat{C}^{ijkl}$\,, 
we only need to look at the $\hh_{00}^2$ terms 
in the reduced Hamiltonian 
\begin{align}
 \mathcal{H} \sim \frac{1}{2} \Bigl[ (\hat{C}^{-1}_{(0)})_{ijkl} \hat{C}^{ijmn}_{(1)}  
 \hat{C}^{klpq}_{(1)} \dot{g}_{mn} \dot{g}_{pq}
 - \hat{C}^{ijkl}_{(2)} \dot{g}_{ij} \dot{g}_{kl} \Bigr]
 - \Bigl[\frac{\hat{N}\sqrt{g}}{2} \,{}^{(3)}\!R \Bigr]_{(2)}
 -\mathcal{L}_m - \mathcal{L}_R\,.
\end{align}
Here, the symbol $\sim$ stands for an equality 
that holds when $\hat{N}^i$ and $\hh_{ij}$ are set to 0. 
$\hat{C}^{ijkl}$ now takes the form 
\begin{align}
 \hat{C}^{ijkl} 
 &=\hat{C}_{(0)}^{ijkl}+\hat{C}_{(1)}^{ijkl}+\hat{C}_{(2)}^{ijkl}+\cdots
\nonumber
\\ &\sim \frac{\sqrt{g}}{4\hat{N}}\, \Bigl[ \frac{1}{2}(g^{ik} g^{jl}
 + g^{il} g^{jk})- g^{ij} g^{kl} \Bigr]
\nonumber
\\
 &= \frac{1}{4} \sqrt{g} \,\Bigl( 1+ \frac{1}{2} \hh_{00}
 +\frac{3}{8} \hh_{00}^2+\cdots \Bigr)
 \Bigl[ \frac{1}{2}(g^{ik} g^{jl}
 + g^{il} g^{jk})- g^{ij} g^{kl} \Bigr]
\end{align}
with
\begin{align}
 \hat{C}_{(0)}^{ijkl} &\sim \frac{\sqrt{g}}{4}\,\Bigl[ \frac{1}{2}(g^{ik} g^{jl}
 + g^{il} g^{jk})- g^{ij} g^{kl} \Bigr]\,,
\\
 \hat{C}_{(1)}^{ijkl} &\sim \frac{1}{2}\,\hh_{00}\,\hat{C}_{(0)}^{ijkl}\,,
\\
 \hat{C}_{(2)}^{ijkl} &\sim \frac{3}{8}\,\hh_{00}^2\,\hat{C}_{(0)}^{ijkl}\,.
\end{align}
Because $\mathcal{L}_m$ does not include $\hh_{00}^2$ terms, 
we thus get 
\begin{align}
 \mathcal{H} &\sim \frac{1}{2}\,\Bigl(\frac{1}{4} - \frac{3}{8} \Bigr)
 \hh_{00}^2\, \hat{C}_{(0)}^{ijkl} \dot{g}_{ij} \dot{g}_{kl} 
 + \frac{\sqrt{g}}{16}\, {}^{(3)}\!R \,\hh_{00}^2 - \mathcal{L}_R 
\nonumber
\\
 &\sim \frac{1}{64} \sqrt{g}\, \bigl[ \dot{g}_{ij} \dot{g}^{ij}
 + (g^{ij}\dot{g}_{ij})^2 \bigr]\,\hh_{00}^2
 + \frac{\sqrt{g}}{16} \,{}^{(3)}\!R\, \hh_{00}^2  -\mathcal{L}_{R}\,.
\end{align}
Finally, we substitute $\hh_{\mu \nu}=2 h_{\mu \nu}$:
\begin{align}
 \mathcal{H} &\sim  \frac{1}{16} \sqrt{g}\, \bigl[ \dot{g}_{ij} \dot{g}^{ij}
 + (g^{ij}\dot{g}_{ij})^2 \bigr]\,h_{00}^2
 + \frac{\sqrt{g}}{4} \,{}^{(3)}\!R\, h_{00}^2  -\mathcal{L}_{R}\,.
\end{align}
From this expression, 
we see that appropriate curvature terms must be supplied by $\mathcal{L}_R$ 
in order for the $h_{00}^2$ terms to disappear. 
To see that this is actually possible, 
we write down the explicit form of $\mathcal{L}_R$ 
for the background metric (\ref{metric_temporal}). 
Necessary formulae are 
\begin{align}
 R&={}^{(3)}\!R + g^{ij}\ddot{g}_{ij} +\frac{3}{4} \dot{g}_{ij} \dot{g}^{ij}
 +\frac{1}{4} (g^{ij}\dot{g}_{ij})^2\,,
\\
 R_{00} &= -\frac{1}{2}g^{ij}\ddot{g}_{ij} -\frac{1}{4} \dot{g}_{ij} \dot{g}^{ij},
\end{align}
from which the $h_{00}^2$ terms involved in (\ref{R term}) 
are obtained as 
\begin{align}
 &\frac{a_1}{2} R_{\mu \nu \rho \sigma} h^{\mu \rho} h^{\nu \sigma} \sim 0\,,
\\
 &\frac{a_2}{2} R_{\mu \nu} h^{\mu \rho} h^\nu_\rho 
 \sim - \frac{a_2}{2} R_{00} h_{00}^2
 = \frac{a_2}{2} \Bigl[ \frac{1}{2}g^{ij}\ddot{g}_{ij}
 +\frac{1}{4} \dot{g}_{ij} \dot{g}^{ij} \Bigr] h_{00}^2\,,
\\
 &\frac{a_3}{2} R h_{\mu \nu} h^{\mu \nu} 
 \sim \frac{a_3}{2} R h_{00}^2 
 = \frac{a_3}{2} \Bigl[ {}^{(3)}\!R +  g^{ij}\ddot{g}_{ij} 
 +\frac{3}{4} \dot{g}_{ij} \dot{g}^{ij}
 +\frac{1}{4} (g^{ij}\dot{g}_{ij})^2 \Bigr] h_{00}^2\,,
\\
 &\frac{b_1}{2} R h^2 
 \sim \frac{b_1}{2} R h_{00}^2 
 = \frac{b_1}{2} \Bigl[ {}^{(3)}\!R + g^{ij}\ddot{g}_{ij} 
 + \frac{3}{4} \dot{g}_{ij} \dot{g}^{ij} 
 + \frac{1}{4} (g^{ij}\dot{g}_{ij})^2 \Bigr] h_{00}^2\,,
\\
 &b_2 R_{\mu \nu} h^{\mu \nu} h 
 \sim - b_2 R_{00} h_{00}^2 
 = b_2 \Bigl[\frac{1}{2}g^{ij}\ddot{g}_{ij} 
 +\frac{1}{4} \dot{g}_{ij} \dot{g}^{ij} \Bigr] h_{00}^2\,.
\end{align}
The reduced Hamiltonian is then expressed as 
\begin{align}
 \mathcal{H}
 &=\frac{\sqrt{g}}{16} \Bigl{\{} 4\,{}^{(3)} \!R
 +\dot{g}_{ij} \dot{g}^{ij} + (g^{ij}\dot{g}_{ij})^2 
 -2( a_2 + 2b_2)( 2g^{ij}\ddot{g}_{ij} +\dot{g}_{ij} \dot{g}^{ij} )
\nonumber
\\
 & \hspace*{60pt} - 2(a_3 + b_1 ) \bigl[ 4\,{}^{(3)}\!R 
 + 4g^{ij}\ddot{g}_{ij} +3\dot{g}_{ij} \dot{g}^{ij} +(g^{ij}\dot{g}_{ij})^2 \bigr]
 \Bigr{\}}\, h_{00}^2\,, 
\end{align}
and we find that 
the necessary and sufficient conditions 
for the coefficients of four independent terms 
${}^{(3)}\!R$, $g^{ij}\ddot{g}_{ij}$, $\dot{g}_{ij} \dot{g}^{ij}$ 
and $(g^{ij}\dot{g}_{ij})^2$ 
to disappear 
are given by the conditions (\ref{constraint1}) and (\ref{constraint2}). 
They are the conditions we promised to show in the beginning of this section 
so that the action (\ref{non-min S})--(\ref{R term}) describes a massive spin 2 field 
with correct DOF in an arbitrary curved background.

\section{Analysis based on the Lagrangian}
\label{nmFP}

In this section we reproduce the results in the previous section 
directly from the Lagrangian without resort to the ADM decomposition. 
We again set the background metric to the form (\ref{metric_temporal})
by using the diffeomorphism invariance.
Then the FP Lagrangian can be written in the following form, 
by decomposing $h_{\mu \nu}$ and their covariant derivatives 
to the temporal and spatial components 
and by integrating by parts appropriately:
\begin{align}
 \mathcal{L} =  \sqrt{g} \bigg{[}&\frac{1}{2}\,C^{ijkl}\dot{h}_{ij}\dot{h}_{kl}
 +\frac{1}{2}M^{ijkl}h_{ij}h_{kl}+D^{ij}\dot{h}_{ij}h_{00}
 +E^{ij}h_{ij}h_{00}
\nonumber
\\
 & +F^{ijk}\dot{h}_{ij}h_{0k} +G^{ijk}h_{ij}h_{0k}+H^ih_{0i}h_{00}
 +\frac{1}{2}I^{ij}h_{0i}h_{0j} +\frac{1}{2}J(h_{00})^2\bigg{]}\,.
\label{nmlgn}
\end{align}
Here, dots denote derivatives with respect to $t$. 
$C^{ijkl}$ does not include curvatures or spatial-derivative operators. 
$I^{ij}$ does not include spatial-derivative operators 
but may include curvatures (as well as $m^2$). 
Note that the FP kinetic term $\mathcal{L}_{\mathcal{E}}$ 
does not contain terms of the form $\dot{h}_{00}\dot{h}_{ij}$\,. 
Completing the square with respect to $\dot{h}_{ij}$ leads to
\begin{align}
 \mathcal{L} = \sqrt{g}\bigg{[}&\frac{1}{2}\,C^{ijkl}
 \Bigl{(}\dot{h}_{ij}+(C^{-1})_{ijmn}(D^{mn}h_{00}+F^{mnp}h_{0p})\Bigr{)}
 \Bigl{(}\dot{h}_{kl}+(C^{-1})_{klqr}(D^{qr}h_{00}+F^{qrs}h_{0s})\Bigr{)}
\nonumber
\\
 & +\frac{1}{2}\,M^{ijkl}h_{ij}h_{kl}+E^{ij}h_{ij}h_{00}
 +\frac{1}{2}J(h_{00})^2+G^{ijk}h_{ij}h_{0k}+H^ih_{0i}h_{00}
 +\frac{1}{2}I^{ij}h_{0i}h_{0j}
\nonumber\\
 &-\frac{1}{2}\,(C^{-1})_{ijkl}\Bigl{(}D^{ij}h_{00}+F^{ijm}h_{0m}\Bigr{)}
 \Bigl{(}D^{kl}h_{00}+F^{kln}h_{0n}\Bigr{)}\bigg{]}\,.
\label{nmlgnquad}
\end{align}
The condition for this Lagrangian to give the proper constraints 
is, as discussed in the previous section, 
that the terms of the form $h_{00}^2$ or $h_{00}\,h_{0i}$ do not survive 
after the Legendre transformation is made with respect to $\dot{h}_{ij}$\,. 
This is translated in the Lagrangian formalism as the condition 
that the second and third lines of (\ref{nmlgnquad}) 
do not give terms of the form $h_{00}^2$ or $h_{00}\,h_{0i}$\,. 
This condition can be written as
\begin{align}
 J-DC^{-1}D = 0\, , 
\label{00quad}
\\
 H^i-(DC^{-1}F)^i = 0\,. 
\label{0icross}
\end{align}
In the following, 
we directly compute the LHS of (\ref{00quad}) and (\ref{0icross}), 
and show that \eqref{0icross} is always satisfied 
but \eqref{00quad} requires the conditions 
\eqref{constraint1} and \eqref{constraint2}.%
\footnote{
 After the first manuscript was accepted for publication, 
 we found that a similar analysis was made in \cite{Buchbinder:1999ar}.
} 

With the metric (\ref{metric_temporal}), the connections are given by 
\begin{align}
 \gm^0_{00} = \gm^0_{0i} = \gm^i_{00} = 0 \,, \quad
 \gm^0_{ij} = \frac{1}{2}\dot{g}_{ij}\,,\quad
 \gm^i_{0j} = \frac{1}{2}g^{ik}\dot{g}_{kj} \,,
\label{connection}
\end{align}
and $\gm^i_{jk}$ agrees with the connection associated with $g_{ij}$. 
Accordingly, the covariant derivatives take the forms 
\begin{align}
 \nb_0h_{00} &=\dot{h}_{00}\,,
\nonumber
\\
 \nb_ih_{00}&=\delu_ih_{00}-2\gm^j_{i0}h_{0j}\,,
\nonumber
\\
 \nb_0h_{0i}&=\dot{h}_{0i}-\gm^j_{0i}h_{0j}\,,
\nonumber
\\
 \nb_jh_{0i}&=\delu_jh_{0i}-\gm^k_{j0}h_{ki}
 -\gm^0_{ji}h_{00}-\gm^k_{ji}h_{0k}\,,
\nonumber
\\
 \nb_0h_{ij}&=\dot{h}_{ij}-\gm^k_{0i}h_{kj}-\gm^k_{0j}h_{ki}\,,
\nonumber
\\
 \nb_kh_{ij}&=\delu_kh_{ij}-\gm^0_{ki}h_{0j}-\gm^0_{kj}h_{0i}
 -\gm^l_{ki}h_{lj}-\gm^l_{kj}h_{li}\,.
\label{coder}
\end{align}
We now write the FP Lagrangian with non-minimal couplings in the following form:
\begin{align}
 \mathcal{L}=\sqrt{-g}&\bigg{[}-\frac{1}{2}\nb_\lambda 
 h_{\mu \nu}\nb^\lambda h^{\mu \nu}
 +\nb^\mu h_{\mu \nu}\nb_\lambda h^{\lambda \nu}
 -\nb^\mu h_{\mu \nu}\nb^\nu h+\frac{1}{2}\nb_\mu h \nb^\mu h 
\nonumber
\\
 &\ - \frac{m^2}{2}(h_{\mu \nu}h^{\mu \nu}-h^2)
\nonumber\\
 &\ +\frac{\tilde{a}_1}{2}R_{\mu\rho\nu\sigma} h^{\mu \nu}h^{\rho \sigma}
 +\frac{\tilde{a}_2}{2}R^{\mu}_{ \ \lambda}h_{\mu \nu}h^{\lambda \nu}
 +\frac{\tilde{a}_3}{2}R h_{\mu \nu}h^{\mu \nu} 
 +\frac{\tilde{b}_1}{2}R h^2 +\tilde{b}_2R_{\mu \nu}h^{\mu \nu}h\bigg{]} \,,
\label{lgntobedecom}
\end{align}
where the parameters are related with those in the previous section, 
\eqref{R term}, as
\begin{align}
 \tilde{a}_1 &= a_1+2\,,\quad \tilde{a}_2 = a_2 + 2\,, \quad
 \tilde{a}_3 = a_3-1\,,
\nonumber
\\
 \tilde{b}_1 &= b_1+\frac{1}{2}\,,\quad \tilde{b}_2 = b_2-1\,.
\end{align}
By substituting (\ref{coder}) to (\ref{lgntobedecom}), 
the coefficients in (\ref{nmlgn}) are expressed as%
\begin{align}
 C^{ijkl} &=\frac{1}{2}(g^{ik}g^{jl}+g^{il}g^{jk})-g^{ij}g^{kl}\,,
\\
 (C^{-1})_{ijkl} &= \frac{1}{2}(g_{ik}g_{jl}+g_{il}g_{jk})
 -\frac{1}{2}g_{ij}g_{kl}\,,
\\
 D^{ij} &= \frac{1}{2}(g^{ik}\gm^j_{k0}+g^{jk}\gm^i_{k0})
 -g^{ij}g^{kl}\gm^0_{kl}\,,
\\
 F^{ijk}h_{0k} &= 2(g^{ij}g^{kl}-g^{ik}g^{jl})\delu_kh_{0l}+O(\gm^2)\,,
\\
 H^ih_{0i} &= \dot{g}^{ij}\delu_ih_{0j}
 +g^{ij}g^{kl}\dot{g}_{kl}\delu_ih_{0j}+O(\gm^2)\,
\\
 J &= 2g^{ij}g^{kl}(\gm^0_{ik}\gm^0_{jl}-\gm^0_{ij}\gm^0_{kl})
 +\frac{1}{2}g^{ij}\dot{g}_{ij}g^{kl}\gm^0_{kl}
\nonumber
\\
 &~~~+\dot{g}^{ij}\gm^0_{ij}+g^{ij}\dot{\gm}^0_{ij}
 +2\tilde{\alpha}R_{00}+2\tilde{\beta}R \,,
\label{nmcoef}
\end{align}
where
\begin{align}
 \tilde{\alpha} &= -\bigg{(}\frac{\tilde{a}_2}{2}+\tilde{b}_2\bigg{)} 
 = -\bigg{(}\frac{a_2}{2}+b_2\bigg{)}\,,
\\
 \tilde{\beta} &= \frac{\tilde{a}_3}{2}+\frac{\tilde{b}_1}{2}
 = \frac{a_3}{2} + \frac{b_1}{2}-\frac{1}{4}\ .
\end{align}
One can easily check that the condition (\ref{0icross}) 
is automatically satisfied (up to higher-order terms). 
On the other hand, 
the LHS of (\ref{00quad}) can be rewritten to the form
\begin{align}
 J - DC^{-1}D  &= \frac{1}{2}(1-2\tilde{\alpha}+4\tilde{\beta})
 g^{ij}\ddot{g}_{ij}
 +\frac{1}{4}(1-2\tilde{\alpha}+6\tilde{\beta})\dot{g}^{ij}\dot{g}_{ij}
 +\frac{1}{2}\tilde{\beta}(g^{ij}\dot{g}_{ij})^2
 +2\tilde{\beta}\,{^{(3)}\!R} \,,
\end{align}
which vanishes only when $\tilde{\alpha}=1/2$ and $\tilde{\beta}=0$, i.e.,
\begin{align}
 \tilde{a}_2+2\tilde{b}_2 = -1\ &\rightarrow\  a_2+2b_2 = -1\,,
\\
 \tilde{a}_3+\tilde{b}_1 = 0\ &\rightarrow\ a_3+b_1=\frac{1}{2} \,.
\label{paradof}
\end{align}
We thus have reproduced the conditions 
(\ref{constraint1}) and (\ref{constraint2}) 
without using the ADM formalism. 
The procedure in this section is a simpler algorithm, 
and might have some application to the analysis of higher spin theories.

\section{Spin 3 case}
\label{spin3}

In this section 
we discuss a massive spin 3 theory in the general background.

The variables to describe a massive spin 3 field 
consist of a traceful, rank-3 symmetric tensor 
$G_{\mu\nu\lambda}$ and an auxiliary scalar $D$\,. 
Denoting the trace of $G_{\mu\nu\lambda}$ by 
$G_\mu \equiv g^{\nu\lambda}G_{\mu\nu\lambda}$\,, 
the  Lagrangian can be written in the form%
%
\begin{align}
 \mathcal{L} = \mathcal{L}_{\mr{min}}+\mathcal{L}_{R}
\label{s3lgn}
\end{align}
with
\begin{align}
 \mathcal{L}_{\mr{min}} &= \sqrt{-g}\,\Bigl[
  {}-\frac{1}{2}\nb_\mu G_{\nu\lambda\rho}\nb^\mu G^{\nu\lambda\rho}
  +\frac{3}{2}\nb^\alpha G_{\alpha\mu\nu}\nb_\beta G^{\beta\mu\nu} 
  -3\nb^\mu G_{\mu\nu\lambda}\nb^\nu G^\lambda
\nonumber \\
  &~~~~ 
  +\frac{3}{2}\nb_\mu G_\nu\nb^\mu G^\nu+\frac{3}{4}\nb^\mu G_\mu\nb^\nu G_\nu 
  +\frac{1}{4}\partial_\mu D\nabla^\mu D 
\nonumber
\\
  &~~~~ -\frac{m^2}{2}\bigl( G_{\mu\nu\lambda}G^{\mu\nu\lambda}
 - 3 G_\mu G^\mu \bigr) 
 + m^2D^2
 - \frac{m}{2}\nb^\mu G_\mu D
 \Bigr]\,, 
\label{s3lgnmin}
\\
 \mathcal{L}_{R} &= \sqrt{-g}\,\Bigl[
  {}\frac{a}{2}\,R_{\mu\nu\lambda\rho}G^{\mu\lambda\alpha}G^{\nu\rho}_{\,\,\,\,\,\,\alpha}
  +\frac{b_1}{2}R_{\mu\nu}G^{\mu\alpha\beta}G^\nu_{\,\,\,\alpha\beta}
  +b_2R_{\mu\nu}G^{\mu\nu\alpha}G_\alpha
  +\frac{b_3}{2}R_{\mu\nu}G^\mu G^\nu 
\nonumber 
\\
  &~~~~ +\frac{c_1}{2}RG_{\mu\nu\lambda}G^{\mu\nu\lambda}
  +\frac{c_2}{2}RG_\mu G^\mu
  +\frac{c_3}{2}RD^2 \Bigr]\,.
\label{s3lgnR}
\end{align}
We will set the background metric to take the form 
\begin{align}
 ds^2 ={} -dt^2+g_{ij}(t)\,dx^idx^j
\label{s3bg}
\end{align}
and assume that all the fields depend only on time $t$\,.  
This setup greatly reduces the amount of necessary calculation, 
and, as we have observed in the preceding sections, 
should be sufficient for investigating 
how the DOF are removed due to constraints.

The coefficients in (\ref{s3lgnmin}) are determined 
such that only the spatial, traceless 
part of the tensor $G_{\mu\nu\lambda}$
is dynamical 
in the flat Minkowski space. 
To confirm this, 
it is convenient to introduce the following parametrization 
for the temporal components of $G_{\mu\nu\lambda}$ 
in the background metric (\ref{s3bg}): 
\begin{align}
 G_{000} = X + 3F \,,\hspace*{12pt}
 G_{00i} = V_i\,,\hspace*{12pt}
 G_{0ij} = \tilde{G}_{0ij}+\frac{1}{3}g_{ij}F\,,
\label{s3variables}
\end{align}
where $F$ is the trace of $G_{0ij}$, $F = g^{jk}G_{0jk}$, 
and $\tilde{G}_{0ij}$ is the traceless part of $G_{0ij}$\,. 
One can easily show that $\tilde{G}_{0ij}$ 
have a nonvanishing quadratic mass term and no kinetic terms, 
which means that $\tilde{G}_{0ij}$ can be removed from the Lagrangian 
algebraically (and thus are not dynamical variables).  
It is also easy to see for the case of flat Minkowski space, 
that the Legendre transformation 
from $\dot{G}_{ijk}$, $\dot{X}$, $\dot{D}$ 
to their conjugate momenta $P^{ijk}$, $P_X$, $P_D$ 
yields only the linear terms for 
$V_i$ and $F$, 
which means that $V_i$ and $F$ play the role of multiplier fields.

In the flat Minkowski case, 
the multipliers $V_i$ and $F$ actually yield the constraints 
that remove all the DOF except for the spatial, traceless 
part of the tensor $G_{\mu\nu\lambda}$\,. 
To see this, 
we note that the dynamics of $(G_{ijk},\,P^{ijk},\,V_i)$ is totally decoupled 
from that of $(X,\,P_X,\,D,\,P_D,\,F)$ 
in our setup. 
We first discuss the subsystem $(G_{ijk},\,P^{ijk},\,V_i)$. 
The primary and secondary constraints with respect to $V_i$ 
are found to be 
\begin{align}
 \kappa^i_1 &\equiv 3m^2\delta_{jk}G^{ijk} = 0\,,\\
 \kappa^i_2 &\equiv -\frac{3}{4}m^2\delta_{jk}P^{ijk} = 0\,, 
\label{s3flatcnstvct}
\end{align}
which have a nonvanishing Poisson bracket, 
$\{\kappa^i_1\,,\kappa^i_2\} = (15/4)m^4 \neq 0$\,. 
Thus, the multipliers $V_i$ remove the DOF of the trace part of $G_{ijk}$ 
and $P^{ijk}$\,, 
and $V_i$ itself is determined by the equation $\dot{\kappa}^i_2=0$\,. 
As for the subsystem $(X,\,P_X,\,D,\,P_D,\,F)$, 
the multiplier $F$ yields 
four constraints (primary, secondary, tertiary and quaternary), 
which are expressed as 
\begin{align}
 \chi_1 &\equiv 2mP_D+2m^2X = 0\,,
\\
 \chi_2 &\equiv 4m^2P_X+4m^3D = 0\,,
\\
 \chi_3 &\equiv -12 m^3P_D-2 m^4X = 0\,,
\\
 \chi_4 &\equiv -4 m^4P_X-24m^5D = 0\,. 
\label{s3flatcnstscl}
\end{align}
Their Poisson brackets take the form 
$\{\chi_1\,,\chi_2\}=\{\chi_1\,,\chi_3\}=0$\,, 
$\{\chi_1\,,\chi_4\}=-40m^6\neq 0$\,, 
and ${\rm det}\{\chi_a,\chi_b\}\neq 0$ $(a,b=1,\ldots,4)$.
Thus, the multiplier $F$ removes the DOF of $(X,P_X,D,P_D)$, 
and $F$ itself is determined by the equation $\dot{\chi}_4=0$\,.

We now require that the same mechanism also work 
for the background (\ref{s3bg}). 
One can easily show that the quadratic terms 
in $V_i$ and $F$ are given by 
\begin{align}
 \mathcal{H}\bigr|_{V_i,F}^{\rm (quad)} 
 &= \sqrt{g}\,\Bigl[ 
 \Bigl(-\frac{3}{4}g^{ij}\ddot{g}_{ij}
 -\frac{3}{8}\dot{g}^{ij}\dot{g}_{ij}\Bigr)V_kV^k 
 +\Bigl(-\frac{3}{2}g^{ik}\ddot{g}_{kj}
 -\frac{3}{4}\dot{g}^{ik}\dot{g}_{kj} \Bigr)V_iV^j 
\nonumber 
\\
 &~~~~ +\Bigl(\frac{31}{6}g^{ij}\ddot{g}_{ij}
 +\frac{31}{12}\dot{g}^{ij}\dot{g}_{ij}\Bigr)F^2 \Bigr]
 - \mathcal{L}_{R}\bigr|_{V_i,F}^{\rm (quad)} 
\end{align}
with
\begin{align}
 &\mathcal{L}_{R}\bigr|_{V_i,F}^{\rm (quad)} / \sqrt{g}
\nonumber
\\
 &= \Bigl[\frac{1}{2}(b_1+b_2+3c_1+c_2)g^{ij}\ddot{g}_{ij}
  +\frac{1}{8}(2b_1+2b_2+9c_1+3c_2)\dot{g}^{ij}\dot{g}_{ij}
 +\frac{1}{8}(3c_1+c_2)(g^{ij}\dot{g}_{ij})^2\Bigr]V_kV^k 
\nonumber 
\\
 &~~~~ +\Bigl[ \frac{1}{4}(-2a+b_1+b_3)g^{ik}\ddot{g}_{kj}
 +\frac{1}{4}(-a+b_1+b_3)\dot{g}^{ik}\dot{g}_{kj}
 +\frac{1}{8}(b_1+b_3)g^{kl}\dot{g}_{kl}g^{im}\dot{g}_{mj}\Bigr] V_iV^j 
\nonumber 
\\
 &~~~~ +\Bigl[\Bigl(\frac{5}{9}a-\frac{43}{18}b_1
 -\frac{8}{3}b_2-b_3-5c_1-2c_2\Bigr)g^{ij}\ddot{g}_{ij}
\nonumber 
\\
 &~~~~~~~~~{} +\Bigl(\frac{19}{72}a-\frac{11}{9}b_1
 -\frac{7}{6}b_2-\frac{1}{2}b_3-\frac{15}{4}c_1-\frac{3}{2}c_2\Bigr)\dot{g}^{ij}\dot{g}_{ij}\nonumber \\
 &~~~~~~~~~{} +\Bigl(-\frac{a}{72}-\frac{b_1}{36}+\frac{1}{6}b_2-\frac{5}{4}c_1-\frac{1}{2}c_2\Bigr)(g^{ij}\dot{g}_{ij})^2 \Bigr]\,F^2\,.
\label{s3hml}
\end{align}
These quadratic terms must vanish 
in order for the $V_i$ and $F$ to give four primary constraints,%
\footnote{
 The primary and secondary constraints $\chi_1$, $\chi_2$ take the forms  
 \begin{align}
 \chi_1 &= 2mP_D + 4 g^{ij}\dot{g}_{ij} P_X+ \sqrt{g}\,(2m^2-\zeta)X\,,
 \nonumber
 \\
 \chi_2 &= -4mg^{ij}\dot{g}_{ij}P_D + (4m^2-\xi)P_X 
 + \sqrt{g}\,(4m^3+2c_3mR)D
 - \sqrt{g}\,(5m^2g^{ij}\dot{g}_{ij}-\eta)X\,,
 \nonumber
 \end{align}
where $\zeta$, $\xi$ and $\eta$ are functions of the curvature. 
In order for the constraints to give the tertiary and quaternary constraints,  
the Poisson bracket $\{\chi_1\,, \chi_2\}$ must vanish. 
However, apparently this does not hold 
at the next order $m^3\times (R/m^2)$ 
for generic backgrounds. 
A detailed analysis on this issue will be reported elsewhere.}
%
and we find that the parameters in the non-minimal couplings must take 
the following values: 
\begin{align}
 a = 3\,,\hspace*{12pt}
 b_1 = -\frac{30}{37}\,,\quad
 b_2 = -\frac{51}{74}\,,\quad
 b_3 = \frac{30}{37}\,,\quad
 c_1 = \frac{119}{222}\,,\quad
 c_2 = -\frac{119}{74}\,.
\label{s3cnd}
\end{align}

\section{Discussion}
\label{discussion}

In this paper we have obtained the Lagrangian 
that describes a free massive spin 2 or spin 3 particle 
propagating in the general gravitational background 
to the first order in the curvature. 
The Lagrangians contain non-minimal couplings.  
For the spin 2 case, 
the coefficients have three free parameters, 
and, in particular, the coupling constant associated with the Riemann tensor is arbitrary.\par  
Actually, there is a well-known theory of massive spin 2 particles. That is the so-called massive gravity theory \cite{deRham:2010ik}\cite{deRham:2010kj}, whose consistency has been proven based on the analysis of the DOF \cite{Hassan:2011vm}\cite{Hassan:2011tf} (for a review, see  \cite{de Rham:2014}\cite{Hinterbichler:2011tt}). We now discuss its relation to our results.%
\footnote {They have developed the massive gravity theory further to construct a theory called bimetric gravity \cite{Hassan:2011zd}. However, because our purpose is to discuss spin 2 particles in the gravitational background, it is more appropriate to consider its original form.}\par
The massive gravity is a non-linear theory, which has a spin 2 massive field $\hat{g}_{\mu \nu}$ and a fixed {\em reference metric} $f_{\mu \nu}$. Here we will consider a classical solution and the fluctuation around it. In general, the classical solution $g_{\mu \nu}$ is determined after $f_{\mu\nu}$ and an initial condition are specified. However, because we are interested in the fluctuation around the classical solution, it is better to regard $f_{\mu \nu}$ as a function of the classical solution $g_{\mu \nu}$. Then the consistency of the EOM for the fluctuation field is automatically guaranteed due to that of the full non-linear theory. We will see that the quadratic Lagrangian for the fluctuation indeed satisfies the conditions (\ref{constraint1}) and (\ref{constraint2}). However, it has only one free parameter, although the massive gravity theory in general has two free parameters. \par 
The action of massive gravity is given by
\begin{align}\label{mg-action}
S = \int d^4x&\sqrt{-\hat{g}}\bigg{[} \frac{1}{2}\hat{R} - m^2\sum_{n=0}^4\alpha_ne_n(\mathbb{K})\bigg{]} \ ,\\
(\mathbb{K})^\mu_{\ \nu} &\equiv (\sqrt{\hat{g}^{-1}f})^\mu_{\ \nu}-\delta^\mu_{\ \nu}\ .
\end{align}
Here $f_{\mu \nu}$ is the reference metric and not a dynamical variable. $\sqrt{\hat{g}^{-1}f}$ denotes the square root as a matrix: $((\sqrt{\hat{g}^{-1}f})^2)^\mu_{ \  \nu} = \hat{g}^{\mu \lambda}f_{\lambda \nu}$. $e_n(\mathbb{K})$ is the elementary symmetric polynomial of degree $n$ in the eigenvalues of  $\mathbb{K}$. They are represented as follows ($[\mathbb{X}]\equiv \mr{tr}\mathbb{X}$):
\begin{align}
e_0(\mathbb{K})&=1\ ,\notag\\
e_1(\mathbb{K})&=[\mathbb{K}]\ ,\notag\\
e_2(\mathbb{K})&=\frac{1}{2}([\mathbb{K}]^2-[\mathbb{K}^2])\ ,\notag\\
e_3(\mathbb{K})&=\frac{1}{6}([\mathbb{K}]^3-3[\mathbb{K}][\mathbb{K}^2]+2[\mathbb{K}^3])\ ,\notag\\
e_4(\mathbb{K})&=\frac{1}{24}([\mathbb{K}]^4-6[\mathbb{K}]^2[\mathbb{K}^2]+3[\mathbb{K}^2]^2+8[\mathbb{K}][\mathbb{K}^3]-6[\mathbb{K}^4])\ .
\end{align}
Several conditions are imposed on the parameters $\alpha_n\ (n=0,\ \cdots,\ 4)$ in order to satisfy the following requirements. We first set $\hat{g}_{\mu \nu} = g_{\mu \nu}+2h_{\mu \nu}$, and expand the Lagrangian with respect to the fluctuation $h_{\mu \nu}$ around $g_{\mu \nu}$. We then require that the first-order terms in $h_{\mu \nu}$ vanish, and that the second-order terms involving $m^2$ take the same form as the FP mass term in the flat background. A straightforward calculation leads to the conditions $\alpha_1=\alpha_0,\ \alpha_2=\alpha_0-1$, and we find that the reference metric $f_{\mu \nu}$ is expressed by $g_{\mu \nu}$ as
\begin{align}\label{mg-feom}
f_{\mu \nu} = g_{\mu \nu}+\frac{2}{m^2}R_{\mu \nu}-\frac{1}{3m^2}g_{\mu \nu}R+O
\bigg{(} \frac{R^2}{m^4} \bigg{)}\ .
\end{align}
Since $\mathbb{K}$ is of first or higher order both in $h_{\mu \nu}$ and in the curvature, $\alpha_4$ does not contribute to the quadratic Lagrangian.\par
After some calculation, we obtain the Lagrangian for the fluctuation
\begin{align}\label{mg-hlgn}
\mathcal{L}\  =\  \sqrt{-g}&\bigg{[}h_{\mu \nu}\mathcal{E}^{\mu \nu \rho \sigma}h_{\rho \sigma}-\frac{m^2}{2}(h_{\mu \nu}h^{\mu \nu}-h^2)\notag\\
&+\frac{2(\alpha_0-\alpha_3)-5}{2}R^{\mu \nu}h_{\mu \lambda}h_\nu^\lambda+\frac{-4(\alpha_0-\alpha_3)+11}{12}Rh_{\mu \nu}h^{\mu \nu}\notag\\
&+\frac{\alpha_0-\alpha_3-2}{3}Rh^2-(\alpha_0-\alpha_3+2)R^{\mu \nu}h_{\mu \nu}h\bigg{]}\ ,
\end{align}
which has the form of (\ref{non-min S})--(\ref{R term}) with
\begin{align}\label{mg-para}
a_1=0\ ,\hspace*{12pt} a_2 = 2(\alpha_0-\alpha_3)- 5\ ,\hspace*{12pt} a_3 = -\frac{2(\alpha_0-\alpha_3)}{3}+\frac{11}{6}\ ,\notag\\
b_1 = \frac{2(\alpha_0-\alpha_3)}{3}-\frac{4}{3}\ ,\hspace*{12pt} b_2 = -(\alpha_0-\alpha_3)+ 2\ .
\end{align}
The coefficients (\ref{mg-para}) indeed satisfy (\ref{paradof}), but depend only on a single parameter $\alpha_0-\alpha_3$. 
We thus may conclude that the Lagrangian in sections \ref{canonical} and \ref{nmFP}
gives a more general description than the massive gravity theory, at least for the free FP field in weak gravitational backgrounds. \par
In this paper only the spin 2 and 3 cases have been discussed. However, it is natural to expect that massive particles with an arbitrary higher spin should also have nontrivial couplings to the curvatures, which we leave as a future work. Although we have not found a Lorentz covariant way to analyze the DOF, such formalism would help to investigate higher spin fields. \par
There are two concrete examples of higher-spin particles in the curved spacetime. One is string theory, where their couplings to gravity can be determined by the scattering amplitudes. The other is composite particles in a well-defined theory such as hadrons in quantum chromodynamics, where in principle we have a description based on the effective Lagrangian.
It will be interesting to compare them with our results, and it might give a clue to the inevitability of string theory.%
\footnote{
See, e.g., \cite{Buchbinder:1999be}\cite{Buchbinder:1999ar} 
for an early study in this direction.
} 

\section*{Acknowledgments}
The authors thank I.L.~Buchbinder, C.~Deffayet, A.~Deriglazov, 
D.~Francia, C.~Germani, L.~Heisenberg, M.~von Strauss and A.~Waldron 
for valuable comments on the first manuscript. 
This work was partially supported by the MEXT 
(MF: Grant No.\ 16K05321, HK: Grant No.\ 16K05322).

\baselineskip=0.83\normalbaselineskip


\end{document}